\documentclass[letterpaper,12pt]{article}
\usepackage[english]{babel}
\usepackage{amssymb,amsmath}
\usepackage{endnotes}

\begin{document}

\title{Where the {\it it from bit} come from?}
\author{Luigi Foschini\\
\footnotesize INAF -- Osservatorio Astronomico di Brera, Merate (LC) -- Italy\\
\footnotesize emails: \texttt{luigi.foschini@brera.inaf.it}; \texttt{foschini.luigi@gmail.com}
}

\date{\today}

\maketitle

In his 1989 essay, John Archibald Wheeler has tried to answer the eternal question of existence \cite{WHEELER1}. He did it by searching for links between information, physics, and quanta. The main concept emerging from his essay is that {\it ``every physical quantity, every it, derives its ultimate significance from bits, binary yes-or-no indications''} \cite{WHEELER1}. This concept has been summarized in the catchphrase {\it ``it from bit''}. In the Wheeler's essay, it is possible to read several times the echoes of the philosophy of Niels Bohr (see \cite{PETERSEN} for a summary). The Danish physicist has pointed out how the quantum and relativistic physics -- forcing us to abandon the anchor of the visual reference of common sense -- have imposed a greater attention to the language. Bohr did not deny the physical reality, but recognizes that there is always need of a language no matter what a person wants to do. To put it as Carlo Sini \cite{SINI}, language is the first toolbox that man has at hands to analyze the experience. It is not a thought translated into words, because to think is to operate with signs as reminded us by various philosophers from Leonardo da Vinci to Ludwig Wittgenstein.

However, reading the autobiography of Wheeler, it seems that the scientist has intended a vision more materialistic than that of Bohr, in which these bits would be true ``quanta of reality'': {\it ``I suggest that we may never understand this strange thing, the quantum, until we understand how information may underlie reality. Information may not be just what we `learn' about the world. It may be what `makes' the world. An example of the idea of it from bit: when a photon is absorbed, and thereby `measured' -- until its absorption, it had no true reality -- an unsplittable bit of information is added to what we know about the world, `and', at the same time, that bit of information determines the structure of one small part of the world. It `creates' the reality of the time and place of that photon's interaction.''} \cite{WHEELER2}.

After Bohr, E. P. Wigner questioned himself about the unreasonable effectiveness of mathematics in physics, but cannot find an answer \cite{WIGNER}. However, he noted some interesting concepts. A major factor in the development of physics is the invariance, in the sense that a stone falls the same way from the Tower of Pisa as from the Empire State Building. In terms of classical physics, the possibility to neglect an enormous quantity of side effects, has allowed us to see how certain physical principles remain valid regardless of space and time. This invariance led to identify those macroscopic regularities in the physical phenomena that induced to think at the existence of universal laws. What emerged from these essays is the existence of an isomorphism between the logic of the adopted language (mathematics) and physical phenomena. It is something very hard to find, hence the need of scientific research and the joys of discovery (or, rather, invention).

This isomorphism is so generic that it is necessary to adjust (calibrate) it from time to time by establishing the connection conditions with specific physical events. It is at this stage that it is set the coupling between the mathematical symbol and a certain measured physical quantity. A physical law is usually expressed in the form of differential equations and initial conditions fix the application of the law of physics with particular events. With the caveat that small changes in initial conditions, then lead to significant changes in the results, as well expressed by the famous phrase of Edward Lorenz \cite{LORENZ}, {\it ``the flap of a butterfly's wings in Brazil set off a tornado in Texas''}. It is as if everything was neglected to develop the general law (differential equation) had been ``swept under the rug'' of initial conditions and their measurement error.

If in classical physics, it was possible to overlook many side effects because of direct sensory experience and human scale, with the relativity and quantum mechanics that was no more possible. This is especially true in quantum mechanics, where the tiny value of the quantum of action has made it impossible to overlook even the smallest effect. Moreover, already at level of classical physics, statistical mechanics had shown that maybe we were neglecting too many things. The determinism of classical mechanics was replaced by probability and statistics.
Brownian motion indicates the transition from an era in which the experiments were within the reach of the senses of a man in an era when certain causes were ``invisible''. At the epoch of Galileo and Newton, the experimental apparatus was a rock and a tower or an inclined plane. It was not necessary a great scientific description. Already the telescope demanded a theory to explain how it worked, but in any case, seeing the moon larger than the naked eye view constituted already solid foothold for the human mind. Instead, nowadays, even the experimental or observational device itself requires complex theories to explain its functioning. For example, we say that a photon is recorded by a CCD because we apply a certain theory, but we cannot really see a photon or the process of converting to electric charge that occurs in the device. Therefore, we have to be careful with language when defining these physical events outside the direct human sensory experience. 

The evolution of physics and the tools used to make it has also imposed the evolution of language. We can also think like Wheeler that, in analogy with the integrated circuits of a computer, there is an information encoded and stored in natural elementary structures at the root of all. Certainly, it is a useful system to operate in conditions with significant background noise. As Shannon wrote \cite{SHANNON}, the fundamental problem of communication is to reproduce an exact or approximate (because of background noise) message generated in another place. Therefore, if you minimize the symbols used (i.e. only 0/1, yes/no), it is possible to neglect anything that has nothing to do with the binary code. Metaphorically, it was what was done with classical physics: to neglect everything that had nothing to do with the physical law under consideration. However, this affects the material part of the message. Instead, the interpretative part escapes from any encoding. But above all, the key question is how the interpretation is generated? Is it intrinsic to the material support or does it depend on an external agent? To answer it is necessary to understand that the problem of information, exploded with quantum mechanics, is nothing else than a linguistic problem, as already highlighted by Bohr. Therefore, it is necessary and essential to study linguistics to understand certain fundamental knowledge.

As I have already written, there is a physical, engineering, material aspect of the information, but it is only one part of a more complex problem. The spoken word is a sequence of pressure changes produced by the vocal cords of a human being. The printed word can be a stretch of ink on a piece of paper or a spatial sequence of pixels turned on and off on the screen of a computer. But this knowledge tells us {\it nothing} on the use that man has made of these material supports. Poets and writers have written poems and novels that have made us dream; physicists developed increasingly complex theories to try understanding the world in which we live; engineers have used these theories to build a technology that has improved our lives. Between the mere material support and its meaning, there is always a unbridgeable gap. Do not think that even linguistics can explain this gap, but at least it can help to frame the problem correctly (\endnote{This ``gap'' has to do with the problem of understanding what consciousness is, or rather, the ``in-consciousness'' (unconscious). It is matter of psychoanalysis or, better, of ciphermatics, the science of the word.}).

In this regard, it is useful to recall the work of Ferdinand de Saussure \cite{SAUSSURE}: {\it what is natural to mankind is not the faculty of speech, but the ability of constructing a language, i.e. a system of distinct signs corresponding to distinct ideas}. This is a fundamental concept: it should be stressed, repeated, written in large letters, carved in marble. Why, in other words, it tells us that the material support of language is of little importance. So that human beings with problems of phonation can equally communicate by means of gestures; humans with visual impairments can use the sensation of touch. There is an interesting experiment carried out by Rawlinson, who tried to assess how the position of the letters of a word changes the understanding \cite{RAWLINSON}. He found that if the first and last letters are in the right position, the internal order of the letters does not matter. For example, ``to raed is good for hatelh''. This applies to a Latin-based tongues, while for other languages with different symbols and grammar, it is necessary to look for other methods. However, this fact puts us once again faced with the fact that the material support is of little importance.

A spoken word can be heard even by a dog or a cat, but they do not come to build a language. Of course, with exercise and with time, one can instruct the animals to associate certain sounds to certain actions, but nothing else. A simple association, a conditioned reflex imposed by an external agent (man), as like the animal can learn to associate a danger to certain natural events. The animals have a very small capacity to build a language, the bare minimum to survive. The man has something more, he can build a tongue. As noted by Guy Consolmagno, his cat is very smart, but she never had in mind to go look through a telescope \cite{CONSOLMAGNO}. 

Therefore, the {\it it from bit} is created by man. It has to do with creativity \cite{FOSCHINI2}. There is no unique pairing, so that there are different mathematical formulations for the same physical phenomena. Think at the formulations of quantum mechanics of Heisenberg, Schr\"odinger, and Dirac; the quantum electrodynamics Schwinger, Tomonaga, Feynman, and Dyson, or even simply to Newtonian, Lagrangian, and Hamiltonian mechanics. It is worth noting that each formulation is not just completely equivalent to the others. There is equivalence in the description of certain phenomena, but then each formulation opens new windows on reality \cite{FOSCHINI1}.

The tongue, according to de Saussure \cite{SAUSSURE}, is both a social product of the faculty of language and a set of necessary conventions, adopted by the society to allow individuals the exercise of this ability. The basic concept of a tongue is the sign, which is divided into signified (meaning) and signifier (material support). The signifier is arbitrary in the sense that there is no specific motivation in his choice as to the meaning. Sometimes, a symbol is used as signifier, but in this case it is never completely arbitrary (the balance as a symbol of justice cannot be replaced with something at random). However, even the sign as a whole somehow always escapes from the individual or social will. This lack of grip, this effect of imbalance of the relationship between signifier and signified, is a very important element of creativity.

One of the main problems of tongue as a social phenomenon is the reproduction of the signs and concepts. The abstraction was instrumental in the evolution of the language and for the birth of mathematics. Easily reproducible signifiers, as the letters of the Latin or Greek alphabet, are at hands of anyone, thus guaranteeing a certain simplicity of communication. Realize signifiers as the frescoes of the Sistine Chapel is not just at hands of anyone ...
The abstraction highlights the conventionality of the signifier, which in turn is taken to extremes in mathematics. The signifier $\pi$ typically indicates the ratio between the circumference and the diameter of a circle, but can even indicate a pion. If I wish, I could also say that it indicates the magnetic moment of a particle: some turn up their nose, but could not advance any reasonable objection on my choice. Mathematics as a tongue for physics requires a signified, built by experiments and observations, directly associated with the signifier. Instead, the semantic field in the spoken tongue is not limited to a direct association with an object or a concept, but also includes contributions from metaphors, metonymies, oxymorons, and other rhetorical figures. The mathematics is therefore a tongue with a reduced semantic field, the necessary price to pay for the search for those invariants that are the basis of classical physics.

What happens if the semantic field extends outside of experiments and observations? It is when one asks oneself what would happen if the parameters of certain theories would change without taking into account the observational or experimental values. For example, think at the multiverse theory, which is obtained by changing the parameters of the standard model and thinking that the universe we observe is simply a random combination of those parameters choice among the infinite possibilities of the multiverse. It is an operation to enlarge the reduced the semantic field of mathematics equivalent -- to draw a parallel with the spoken tongue -- to the use of rhetorical figures. At this point, a mathematical tongue similar in structure to the spoken tongue allows you to do also works of fantasy. If physics is what we can say about the nature, as written by Bohr, then to wonder what happens by changing the parameters derived from experiments and observations is no more physics, but science fiction (or even pataphysics). This was already done by many writers (and some scientists) with much greater success. Not to mention that ask why the Universe is done in a way and not in another is more a religious issue rather than a scientific one. The will to explain it by means of science comes from the desire of atheism of some, which is incompatible with science. If one wants to persist with atheism, then it is better to think that the universe is the product of pure chance, a fluctuation, rather than to invent multiverse. The maximum of Wittgenstein must be reminded\cite{WITTGENSTEIN}: {\it ``Whereof one cannot speak, thereof one must be silent''}. Or, at least, if you still want to talk, it is good to have a clear idea of the creative power of language.

In summary to conclude, it is now clear that the knowledge of the material aspect only is not sufficient to understand the problem of information. Nature does not operate: even if the material support can evolve and change, this has a secondary importance. It is the human being that, by assigning a meaning and creating a tongue with the signs so obtained, creates the {\it it from bit}. How this happen has to do with the human mind and its conscious and unconscious structures. One can be satisfied for thinking that it is a gift (e.g. Wigner) or can try to understand how this happen. To do this we need to equip ourselves with the necessary knowledge, expanding our own culture to ``hybrid'' dimensions (\endnote{{\tt http://manifestoibridi.org/}}).
Without doubt, it is a problem of fundamental importance for any field of science, not just physics. The understanding that there are linguistic and psychic issues, helps us to correctly contextualize the problem, avoiding to hunt for ghosts. For example, the question asked by FQXi ``How does the nature (the universe and the things therein) `store' and `process' information?'' does not make sense.

\vskip 12pt
\emph{I thank S. Dalla Val for interesting discussions of linguistics and a critical reading of the text.}

\theendnotes

\end{document}